\documentclass[aps,prx,twocolumn,amssymb,superscriptaddress]{revtex4-2}
\usepackage{subfigure}
\usepackage{graphicx}
\usepackage{rotating} 
\usepackage{color}
\usepackage{rotating}
 \usepackage{amsmath,amsthm}
 \usepackage{enumitem}
\usepackage{epstopdf}
\begin{document}
\title{Solvent quality dependent osmotic pressure of polymer solutions in two dimensions}
\author{Lei Liu}
\thanks{leiliu@zstu.edu.cn}
\affiliation{Key Laboratory of Optical Field Manipulation of Zhejiang Province, Department of Physics, Zhejiang Sci-Tech University, Hangzhou 310018, China}
\author{Changbong Hyeon}
\thanks{hyeoncb@kias.re.kr}
\affiliation{Korea Institute for Advanced Study, Seoul 02455, Korea}
\date{\today}

\begin{abstract}
Confined in two dimensional planes, polymer chains comprising dense monolayer solution are segregated from each other due to topological interaction. 
Although the segregation is inherent in two dimensions (2D), 
the solution may display different properties depending on the solvent quality. 
Among others, it is well known {\color{black}in both theory and experiment} that the osmotic pressure ($\Pi$) in the semi-dilute regime displays solvent quality-dependent increases with the area fraction ($\phi$) (or monomer concentration, $\rho$),   
that is, $\Pi\sim \phi^3$ for good solvent and $\Pi\sim \phi^8$ for $\Theta$ solvent. 
The osmotic pressure can be associated with the Flory exponent (or the correlation length exponent) for the chain size and the pair distribution function of monomers; however, they do not necessarily offer a detailed microscopic picture leading to the difference. 
{\color{black}To gain microscopic understanding into the different surface pressure isotherms of polymer solution under the two distinct solvent conditions, 
we study the chain configurations of polymer solution based on our numerical simulations that semi-quantitatively reproduce the expected scaling behaviors.} 
Notably, at the same value of $\phi$, polymer chains in $\Theta$ solvent occupy the surface in a more \emph{inhomogeneous} manner than the chains in good solvent, yielding on average a greater and more heterogeneous interstitial void size, which is related to the fact that the polymer in $\Theta$ solvent has a greater correlation length. 
The polymer configurations and interstitial voids visualized and quantitatively analyzed in this study offer microscopic understanding to the origin of the solvent quality dependent osmotic pressure of 2D 
polymer solutions. 
 \end{abstract}
\maketitle 

\section{Introduction}
A flexible polymer chain in three dimensions (3D) exhibits a coil-to-globule transition with varying temperature ($T$) or solvent quality. 
{\color{black}
The polymer size, so called Flory radius $R_F$, scales with the number of monomers ($N$) as $R_F\sim N^{\nu}$. 
The scaling exponent $\nu$, known as the Flory exponent, which is tantamount to the correlation length exponent in the general context of critical phenomena \cite{deGennesbook}, changes from $\nu=0.588$ (good, $T>\Theta$) to $\nu=1/3$ (poor, $T<\Theta$) as the temperature is lowered.} 
At $T\approx \Theta$, the exponent $\nu_{\Theta}^{3D}=1/2$ is identical to that of the random walk (RW). 
The polymer chains in $\Theta$ solvent are at a point where the attraction and repulsion between monomers compensate. 
The $\Theta$ point of 3D polymer is determined at which the second virial coefficient vanishes ($B_2=0$). 
The higher-order virial terms contribute only logarithmically to the free energy, so that the condition of $N\gg 1$ yields $R_F\sim N^{1/2}$ \cite{Duplantier87JCP}. 
In light of the polymer-magnet analogy, the condition $B_2=0$ amounts to the tricritical point of the Landau free energy \cite{deGennes75JPL}. 

In two dimensions (2D), the correlation length exponent is 
$\nu_{\Theta}^{2D}=4/7$, which is different from that of RW.  
{\color{black}Since the higher-order virial terms cannot be ignored in 2D, }
the $\Theta$ point of polymer chain in 2D is defined under a more subtle condition than $B_2=0$ in 3D \cite{douglas1985polymers,Liu19Nanolett,Jung2020Macromol}. 
{\color{black}As a result, the $\Theta$ point of 2D polymer, in practice, has numerically been} attained by tuning the relevant parameters~\cite{coniglio1987PRB} (see Methods).     
Historically, studies on geometrical fractal objects in 2D, in particular, the $\Theta$ chain in 2D and its exotic exponent $\nu_{\Theta}^{2D}=4/7\approx 0.571$ culminated in 1980s.  
Coniglio \emph{et al.} \cite{coniglio1987PRB} posited that the fractal dimension ($\mathcal{D}=(\nu_{\Theta}^{2D})^{-1}=7/4$) of the 2D \emph{interacting} self-avoiding walk at the coil-to-globule transition point is identical to the dimension of percolating clusters' boundaries (hulls)~\cite{Sapoval85JPL,Bunde85JPA}. 
Duplantier and Saleur showed using the conformal invariance that the 2D $\Theta$ chain, 
the hull of percolating clusters, and uncorrelated diffusion fronts in 2D at scaling limit are all characterized with the fractal (Hausdorff) dimension of $7/4$ and belong to the same universality class \cite{Duplantier87PRL,saleur1987PRL,duplantier1989PRL,duplantier1989PR}.  
Later this result was more rigorously proven as the Hausdorff dimension of the curve ($\mathcal{D}=\min{\left(2,1+\kappa/8\right)}$) generated from Stochastic-Loewner Evolution (SLE) process with parameter $\kappa=6$, denoted by SLE$_{6}$ \cite{beffara2004hausdorff}. 

\begin{figure*}[ht!]
        \includegraphics[width=0.9\linewidth]{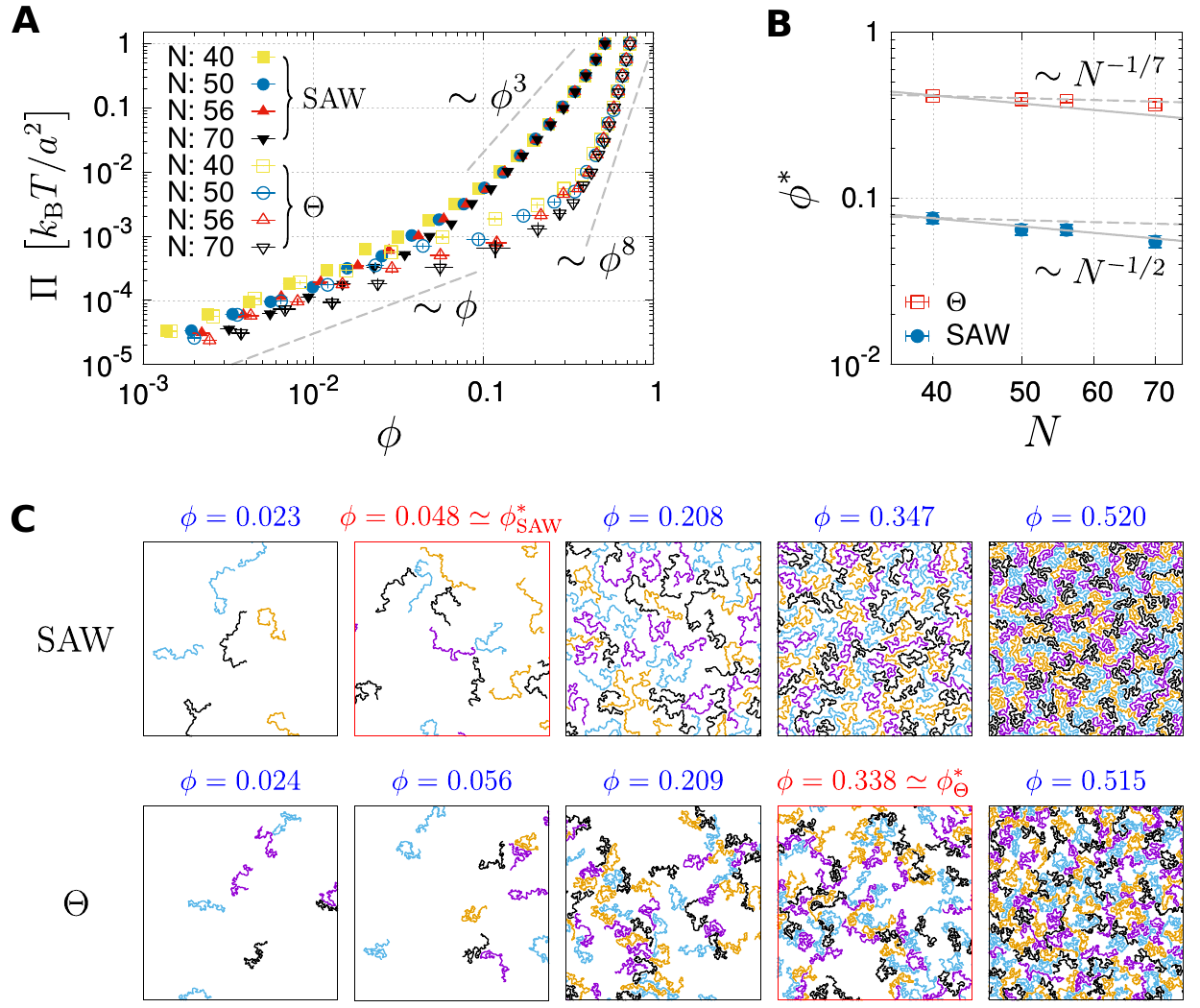}
		\caption{Difference between 2D polymer solutions made of SAW and $\Theta$ chains. (A) $\Pi$ versus $\phi$ with varying $N$. 
		(B) $N$-dependent overlap fraction ($\phi^\ast$) from dilute to semi-dilute solution. 
		$\phi^\ast$ for each $N$ is determined from the crossover point where the fits using $\phi$ and $\phi^3$ (or $\phi$ and $\phi^8$) lines meet in (A). 
		For $N=70$, $\phi^\ast_\text{SAW}=0.054\pm 0.004$ and $\phi^\ast_\Theta=0.365 \pm 0.015$.  		
		Dashed and solid lines are the $N$-dependences expected for $\phi^\ast_\Theta\sim N^{-1/7}$ and $\phi^\ast_\text{SAW}\sim N^{-1/2}$.  
     (C) From left to right, shown are the SAW (top row) and $\Theta$ polymer solutions (bottom row) with increasing $\phi$, encompassing the non-overlapping ($\phi/\phi^*<1$), semi-dilute ($\phi/\phi^*\approx 1$), and dense melt regimes ($\phi/\phi^*\gg 1$). 
The panels corresponding to the area fraction close to $\phi^\ast$ are enclosed in the red boxes.  
		Each chain is shown in different colors. 
All the configurations of solutions consisting of monodisperse polymers with $N=70$ are shown in the 2D box of the same size. 
As a result, only a part of the simulation is depicted except for the case with the highest $\phi$. See Figs.~\ref{conf_S1} and S1 for the configurations of polymer solution with $N=640$. 
		\label{conf1}}
\end{figure*}

\begin{figure*}[ht!]
\includegraphics[width=1\linewidth]{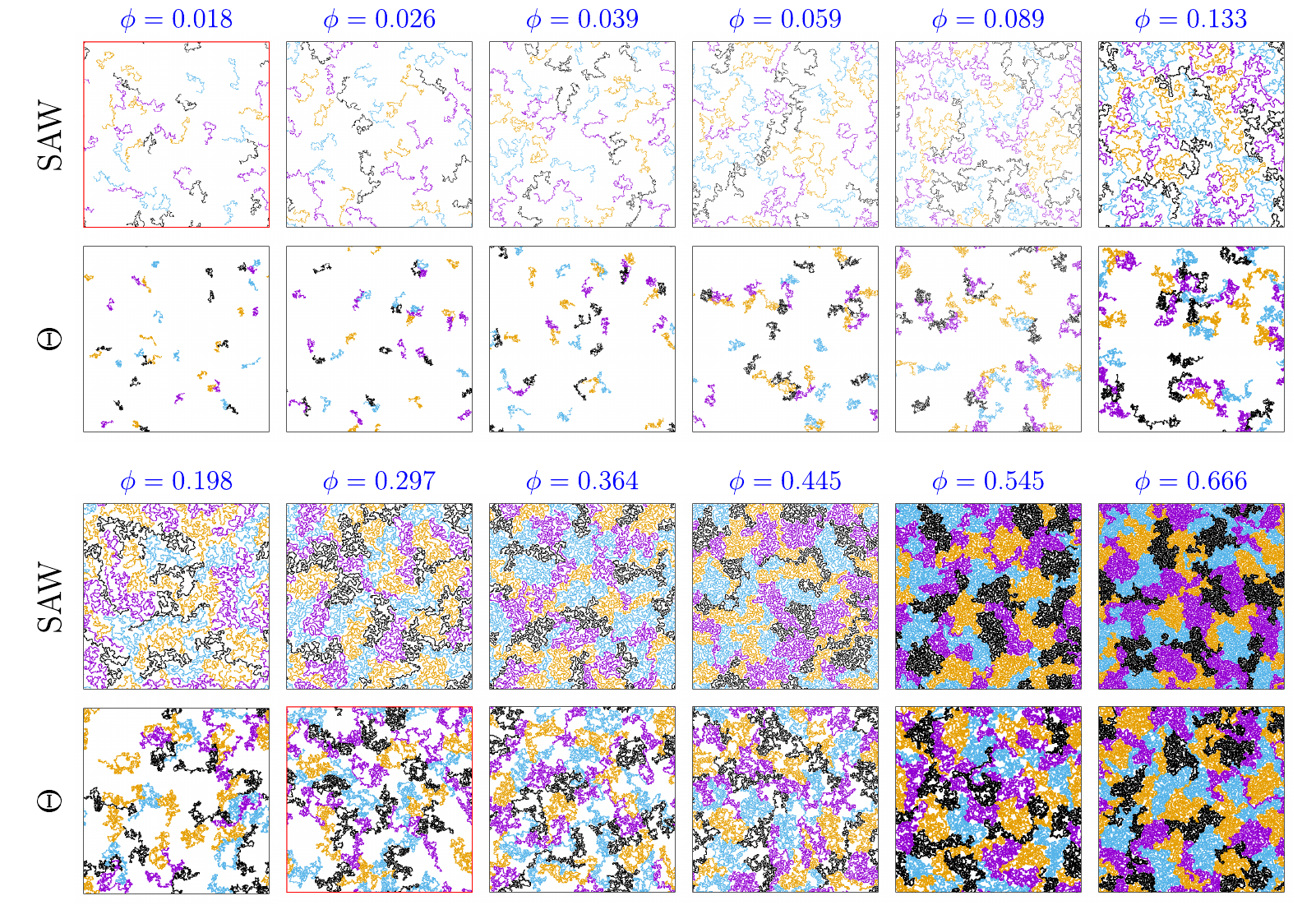}
\caption{
Polymer solutions of SAW and $\Theta$ chains in 2D with increasing area fraction $\phi$ (or equivalently the monomer density), 
from non-overlapping ($\phi/\phi^*<1$), to semi-dilute ($\phi/\phi^*\approx 1$), and dense melt regimes ($\phi/\phi^*\gg 1$). 
Each chain with $N=640$ is depicted in different colors.  
The threshold overlap area fractions for SAW and $\Theta$ solutions with $N=640$ are estimated by 
extrapolating the relation $\phi^\ast_\text{SAW}\sim N^{-1/2}$ and $\phi^\ast_\Theta\sim N^{-1/7}$ and the knowledge of $\phi^\ast$ at $N=70$ (Fig.~\ref{conf1}B) to $N=640$, which yields $\phi^\ast_\text{SAW}\approx 0.018$ and $\phi^\ast_\Theta\approx 0.266$ for $N=640$. 
Note that each panel is drawn by maintaining the relation of $\phi\times L^2=const$.  
In other words, the number of polymers (monomers) in each panel is identical, and that a panel with smaller $\phi$ 
displays a larger simulation box size ($L\times L$). 
Fig.~S1 offers the snapshots of the simulations conducted at different $\phi$ in the 2D box at the identical field of view. 
}
\label{conf_S1}
\end{figure*}

The correlation length exponent $\nu$ of 2D polymer has indirectly been determined through surface pressure ($\Pi$) measurements of thin polymer film formed at air-water interface as a function of the area fraction of polymer solution ($\phi$) \cite{vilanove1980PRL,witte2010macromolecules}. 
In the semi-dilute phase $\phi^*<\phi \ll 1$, where $\phi^*$ is the critical overlap area fraction. 
(See Appendix A for the basics of osmotic pressure of polymer solutions with increasing $\phi$), 
$\Pi$ scales with $\phi$ as $\Pi \sim \phi^q$. 
Actual measurements of the surface pressure (or osmotic pressure) have shown that  
$q=3$ for 2D polymer solutions in good solvent, 
whereas  $q=8$ in $\Theta$ solvent~\cite{vilanove1980PRL,cicuta2004scaling,witte2010macromolecules}. 
Since the exponent $q$ is related with $\nu$ as $q=2\nu/(2\nu-1)$~\cite{deGennesbook} (see Appendix A), 
it can be deduced from $\Pi\sim \phi^{q}$ that a single polymer chain in 2D obeys $R_F\sim N^{\nu}$ with $\nu=3/4=0.75$ and $\nu=4/7\approx 0.571$ under good and $\Theta$ solvent conditions, respectively. 
The difference between the exponents ($q$) of the osmotic pressure against $\phi$ in the semi-dilute phase under the two solvent conditions is significant. 
However, 
besides the difference in $\nu$, 
it remains elusive how the $\phi$-dependent configurations of individual polymer chains and their interface with neighboring chains contribute to the surface pressure under 
the two different solvent conditions.

Despite a number of extensive theoretical studies on the thermodynamics and conformational properties of 2D polymer solutions, their focus was predominantly on the solution made of self-avoiding polymers  \cite{carmesin1990JP,nelson1997JCP,wang2000macromolecules,yethiraj2003Macromolecules,semenov2003EPJE,sung2010PRE,schulmann2013PSSC}. 
Here, using theoretical arguments along with numerics, 
we aim to explore the microscopic underpinning that leads to the solvent quality dependent $\Pi$-$\phi$ isotherm of 2D polymer solutions. 
In this paper, 
we first demonstrate the main result of $\Pi$-$\phi$ isotherm calculated from the simulations of polymer solutions {\color{black}comprised of self-avoiding walks (SAWs) and $\Theta$ chains.} 
We next discuss the difference of two polymer solutions in terms of the radial distribution of monomers ($g(r)$), 
solvent quality dependent conformations of individual polymer chains, and interstitial voids formed in  polymer solutions. 
{\color{black}Our study highlights  
the distinct chain statistics and organization of polymer solution under good and $\Theta$ solvent conditions, offering a clear picture of how these differences 
lead to distinct $\Pi$-$\phi$ isotherms of polymer solution in 2D.}

\begin{figure}[ht!]
\includegraphics[width=1.0\linewidth]{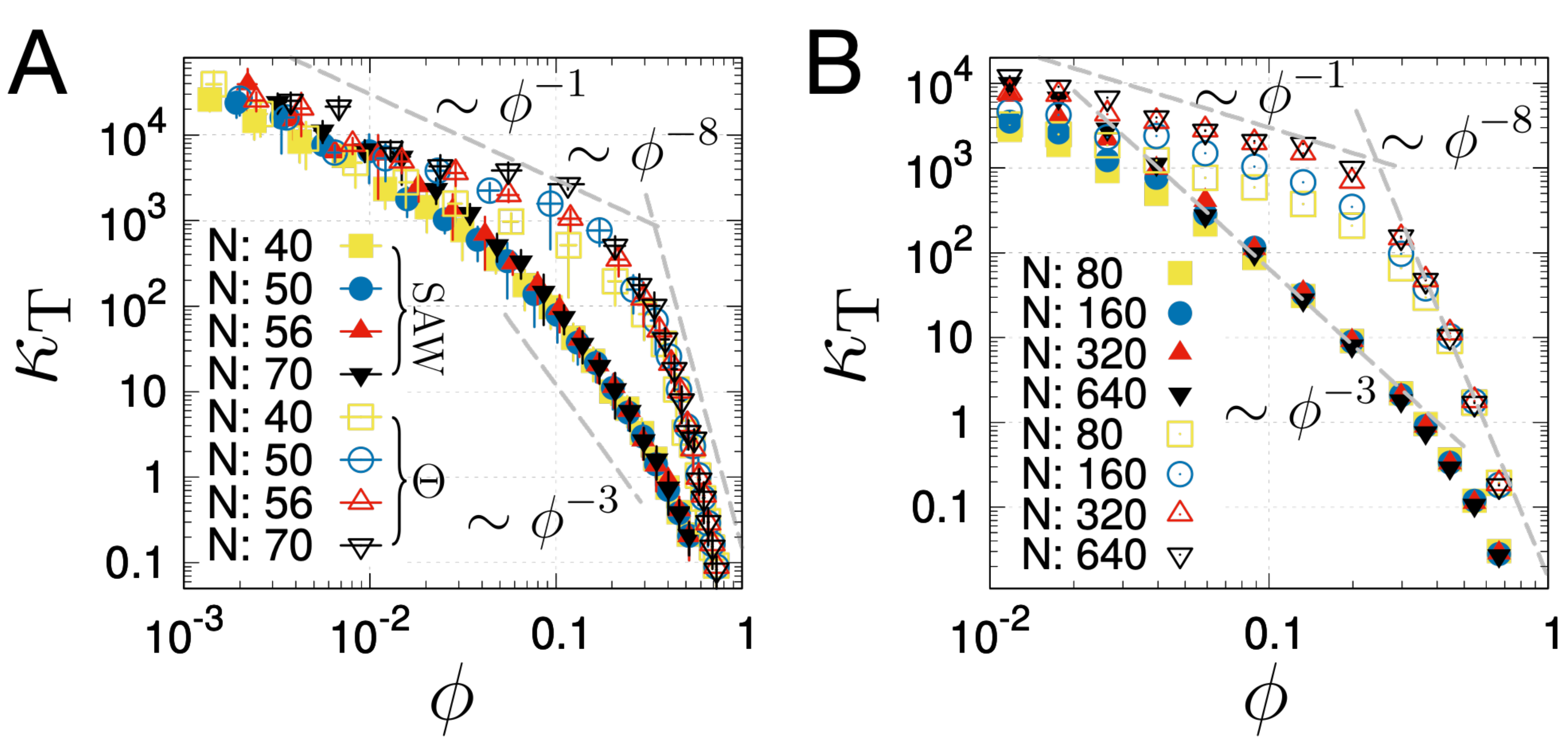}
\caption{
Standard isothermal compressibility $\kappa_\text{T}$ versus the area fraction. 
(A) For short chains, simulations were performed in an isothermal-isobaric (NPT) ensemble. 
The compressibility was calculated by $\kappa_\text{T} = \left(\left<A^2\right> - \left<A\right>^{2}\right)/\left<A\right> k_\text{B}T$ where $A$ is the fluctuating area of the simulated solution \cite{Hansen1990Book}. 
(B) For longer chains, $\kappa_\text{T}$ was calculated from the particle number fluctuations in subdomains in a canonical ensemble by using a spatial block analysis method \cite{heidari2018Entropy}. 
More details are explained in Fig.~S3 and its caption. 
Since the dimensionless reduced isothermal compressibility $\chi_\text{T}$ ($= \rho k_\text{B}T \kappa_\text{T}$), 
defined by the ratio between bulk isothermal compressibility and that of the ideal gas $\left(\rho k_{B}T\right)^{-1}$, is expected to scale with the blob size $g(\rho)$ \cite{schulmann2013PSSC}, 
$\kappa_\text{T}$ scales as $\phi^{-3}$ and $\phi^{-8}$ in semi-dilute SAW and $\Theta$ chain solutions, respectively. 
}
\label{kappaT}
\end{figure}

\section{Results} 
\subsection*{Pressure isotherm of polymer solution confined in two dimensions} 
Fig.~\ref{conf1}A demonstrates the osmotic pressure calculated for 2D polymer solution 
consisting of monodisperse chains with varying $\phi$ and $N$ ($N=40$, 50, 56, and 70) under good and $\Theta$ solvent conditions, {\color{black}with the chain configurations with $N=70$ shown in Fig.~\ref{conf1}C (see Fig.~\ref{conf_S1} for the chain configurations with $N=640$ for increasing $\phi$).} 
The polymer solution were simulated on 2D plane of area $A(=L^2)$ along with periodic boundary condition. 
The osmotic pressure was computed by evaluating the virial equation \cite{schulmann2013PSSC} 
\begin{align}
\Pi=\rho k_BT-\frac{1}{2A}\sum_{i<j}r_{ij}\frac{du(r)}{dr}\Big|_{r=r_{ij}}, 
\end{align}
where $u(r)$ (see Eq.~\ref{eqn:potential}) is the inter-particle potential between two monomers separated by a distance $r$. 

The $\Pi$-$\phi$ isotherm calculated from the simulations {\color{black}semi-quantitatively} reproduces the theoretically anticipated scaling behaviors (see Appendix A).  
(i) For $\phi<\phi^\ast$ the osmotic pressure $\Pi$ of the 2D polymer solution scales linearly with the area fraction $\phi$ as $\Pi\sim \phi/N$. 
Note that {\color{black}for the same $\phi$} $\Pi$ is indeed smaller for a larger chain length $N$ (see Fig.~\ref{conf1}A). 
(ii) For $\phi^\ast < \phi\ll 1$, $\Pi_\text{SAW}\sim \phi^3$, 
$\Pi_\Theta\sim \phi^8$, and $\Pi$-$\phi$ isotherm becomes independent of $N$ (see Fig.~\ref{conf1}A and Appendix A).   
(iii) In all values of $\phi$, we find that $\Pi_{\rm SAW}(\phi)>\Pi_{\Theta}(\phi)$. 
(iv) The overlap area fraction $\phi^\ast$ decreases with $N$ as $\phi^\ast\sim N^{1-2\nu}$. 
Using the scaling relations $\phi^\ast_\Theta\sim N^{-1/7}$ and $\phi^\ast_\text{SAW}\sim N^{-1/2}$, one can extrapolate $\phi^\ast$ for large $N$ (see Fig.~\ref{conf1}B).

\begin{figure}[t]
\includegraphics[width=1.0\linewidth]{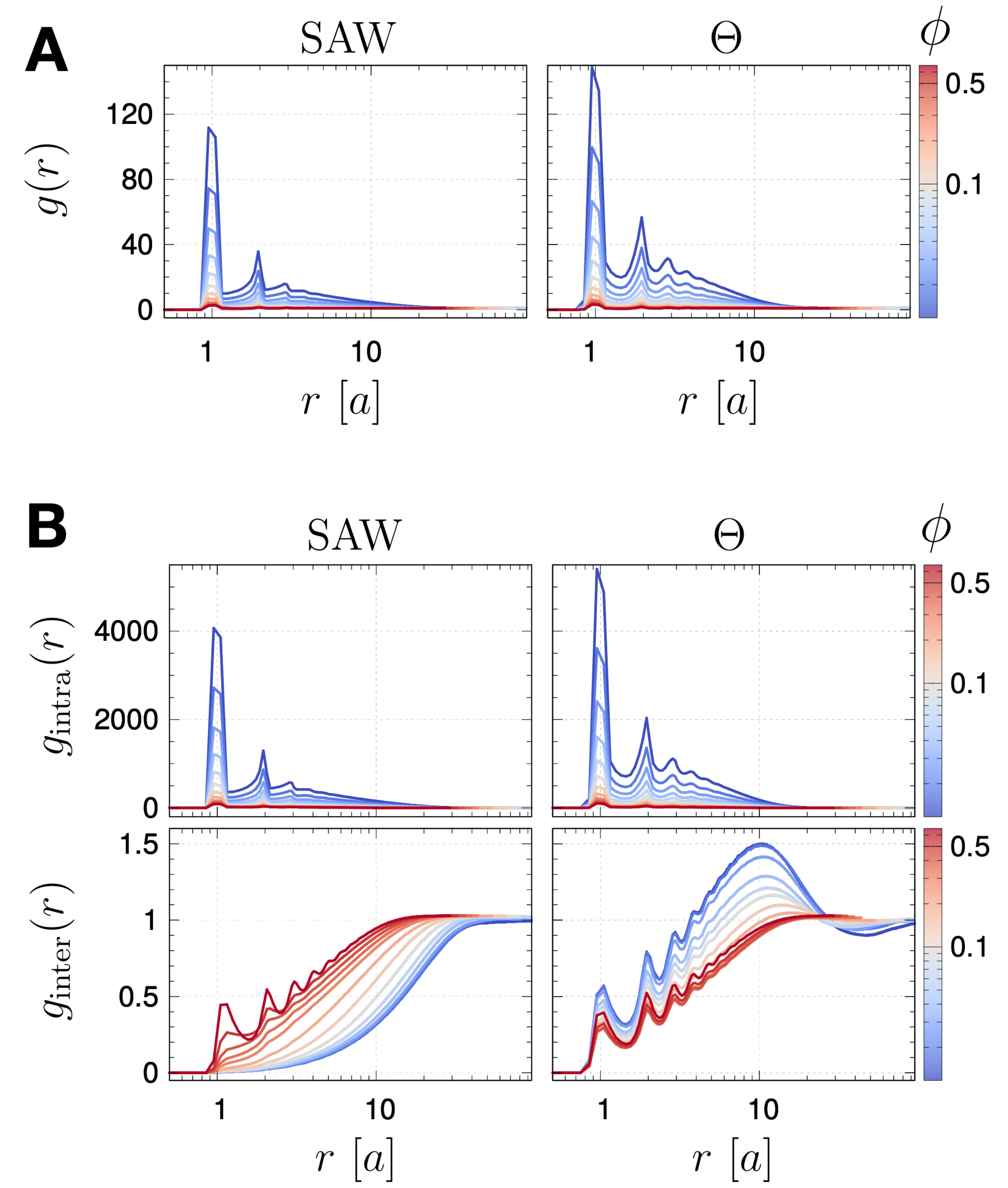}
\caption{
The radial distribution of monomers ($g(r)$) in SAW (left) and $\Theta$ polymer solutions (right) at varying $\phi$. 
The distribution was calculated using polymer solutions with $N=70$. 
(A) The full radial distribution of monomers. 
(B) The full radial distribution of monomers decomposed into the intra- (top) and inter-chain (bottom) radial distributions. 
}
\label{fig:gr}
\end{figure}

\begin{figure}[t]
\includegraphics[width=1.0\linewidth]{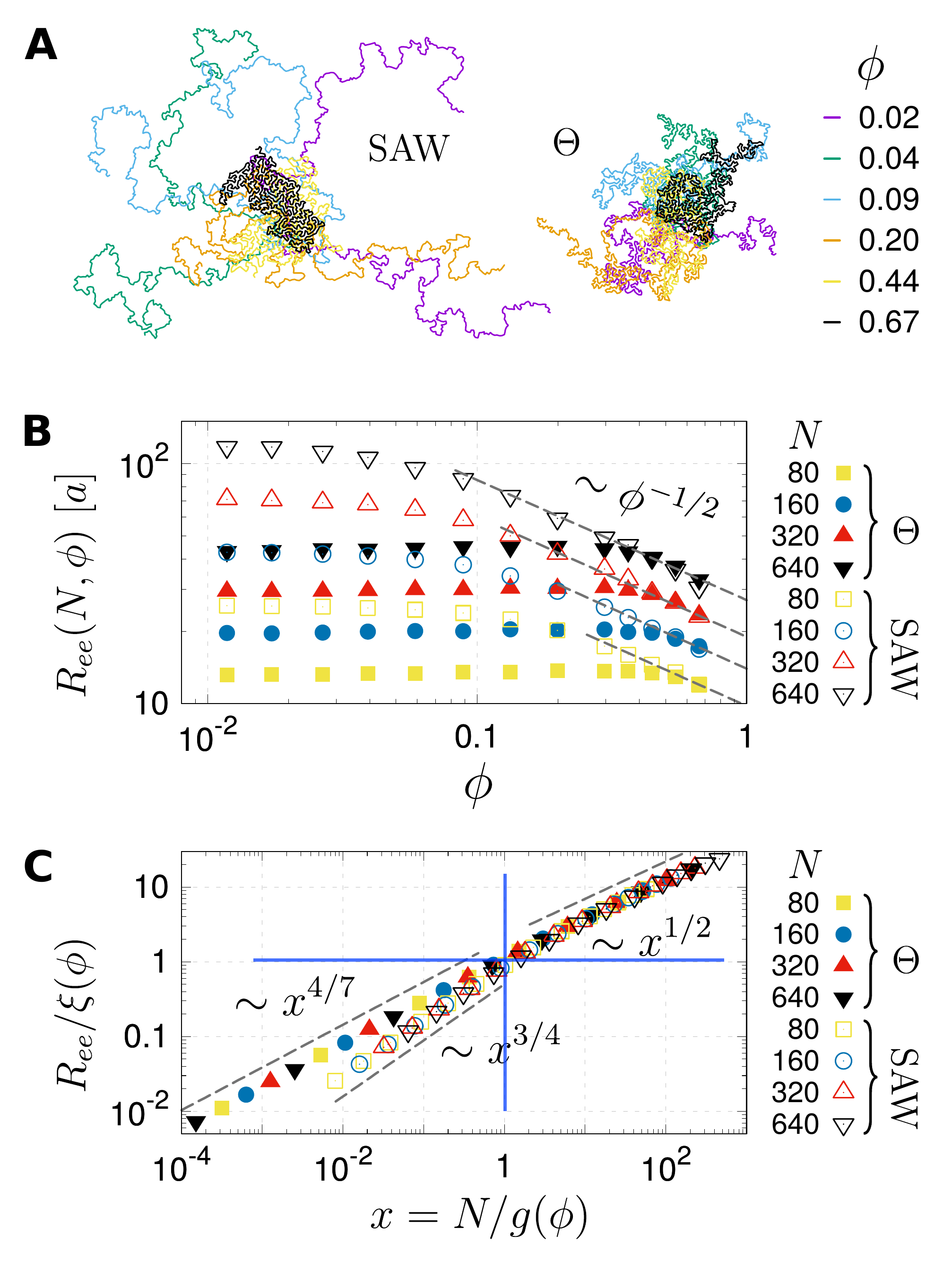}
\caption{
(A) Configurations of a polymer chain ($N=640$) in 2D polymer solution in good solvent and $\Theta$ condition. 
The figure shows how the size of a single polymer changes with varying $\phi$. 
At the same value of $\phi$, polymer chains in $\Theta$ condition are generally more compact that those in good solvent condition. 
(B) The size of individual chain in terms of the end-to-end distance ($R_{ee}$) versus $\phi$. 
At high $\phi$, both polymers under good and $\Theta$ solvent conditions form dense polymer melts and their size scales with $\phi$ as $R_{ee}\sim \phi^{-1/2}$. 
(C) The chain size (end-to-end distance, $R_{ee}$) rescaled with a blob size ($\xi\simeq a\phi^{\frac{\nu}{1-2\nu}}$) plotted against $N$ rescaled with the number of monomers in a blob ($g\simeq \phi^{\frac{1}{1-2\nu}}$) for varying densities collapse to the master curves for SAW and $\Theta$ solution.  
In both cases, the cross-over points to dense polymer melts are identified at $N/g(\phi)\simeq 1$ and $R_{ee}/\xi(\phi)\simeq 1$. 
}
\label{singConf}
\end{figure}

\section{Discussions}
\subsection*{The radial distribution of monomers determines the scaling behavior of osmotic pressure}
The compressibility equation from the theory of liquids \cite{Hansen1990Book} associates  
the number fluctuations ($\langle(\delta N)^2\rangle$) with the isothermal compressibility ($\kappa_T$) and the radial distribution function ($g(r)$) as follows (see Appendix B for the details of derivation). 
\begin{align}
\frac{\langle(\delta N)^2\rangle}{\langle N\rangle}=\rho k_BT \kappa_T=1+\rho\int \left(g({\bf r})-1\right)d{\bf r}. 
\label{eqn:gr_kappa}
\end{align}
From the definition of isothermal compressibility, 
\begin{align}
\kappa_T=-\frac{1}{A}\left(\frac{\partial A}{\partial \Pi}\right)_{T,N}=\frac{1}{\phi}\left(\frac{\partial \phi}{\partial \Pi}\right)_{T,N}
\end{align}
where $\phi\simeq Na^2/A$, it is straightforward to show that 
the scaling relationship of $\Pi\sim \phi^{q}$ signifies $\kappa_T\sim\phi^{-q}$, which is confirmed explicitly in {\color{black}Fig.~\ref{kappaT}}. 

According to Eq.~\ref{eqn:gr_kappa}, 
the solvent quality-dependent $\Pi$-$\phi$ isotherm originates from the difference in $g(r)$. Thus, there ought to be difference between $g(r)$'s of SAW and $\Theta$ polymer solution, and the difference should yield the distinct exponent $q$. 
Nevertheless, it is not straightforward to see 
the difference between the $g(r)$'s of the two solvent conditions except for the amplitude ({\color{black}Fig.~\ref{fig:gr}A}). 
Only when the $g(r)$ is decomposed into the contributions from the particles comprising the same chain ($g_\text{intra}(r)$) and different chain ($g_\text{inter}(r)$) ({\color{black}Fig.~\ref{fig:gr}B}) 
(or see their Fourier transformed version of $g_{\rm intra}(r)$, $F(k)$, calculated in Fig.~S2), 
it becomes clear that there is a qualitative difference between $g^{\rm SAW}_{\rm inter}(r)$ and $g^{\Theta}_{\rm inter}(r)$; however such a decomposition is of limited use in that it is not directly accessible in experimental measurements.

\subsection*{Chain conformations in polymer solution} 
The conformations of polymer chains with increasing $\phi$ illustrated in Fig.~\ref{conf1}C draw a distinction between the two types of polymer solution. 
Better visualization of the chain conformations in 2D polymer solutions calculated with 
longer polymer chains with $N=640$ is given in {\color{black}Figs.~\ref{conf_S1} and S2}.  

First, the individual chains in $\Theta$ solvent are more crumpled than those in good solvent. 
Over the intermediate regime of the wave vector $k$ ($1/R_g<k<1/\xi(\phi)$), corresponding to the length scale of $6a<r<60a$, the structure factor $F(k)$ scales with $k$ as 
$F(k)\sim k^{-4/3}$ and $F(k)\sim k^{-7/4}$ for polymer solutions under good and $\Theta$ solvent conditions, respectively (see Fig.~S2), reflecting the difference between the spatial arrangement of the monomers in the two polymer solutions.  

In the semi-dilute phase ($\phi > \phi^\ast$), 
with increasing $\phi$, the size of individual chain (or domain size) decreases for the case of SAW solution; however, such tendency is effectively absent in $\Theta$ polymer solution for $0.02\leq\phi\leq 0.44$ (see Fig.~\ref{singConf}A). 
The snapshots of individual chains from the simulations at varying $\phi$  shown in Fig.~\ref{singConf}A 
indicate that compared to the chains in $\Theta$ solvent, the extent of size reduction in the chain under good solvent condition is greater. 
The size of $\Theta$ chain is not sensitive to $\phi$ as if each chain barely feels the neighboring chains. 
The insensitivity of $\Theta$ polymer size to the increasing $\phi$ can also be confirmed with the intrachain form factor ($F(k)$) calculated in Fig.~S2. 

To be more quantitative, 
we calculate the mean end-to-end distance of individual chains as a function of $\phi$ (Fig.~\ref{singConf}B).  
Two points are noteworthy. 
(i) As long as the solution is not in the concentrated regime ($\phi\ll 1$), $\Theta$ chains in solution (filled symbols in Fig.~\ref{singConf}B) maintain their size. The minor expansions observed at $\phi < \phi^\ast$ is likely due to the effect of the neighboring chains that attract. 
(ii) At sufficiently high area fraction ($\phi\gg \phi^\ast$), the polymer solution becomes a melt, 
the data points with the same $N$ from the two solvent conditions coincide, and the effect of solvent quality on polymer size is no longer observed. 
The sizes of chains are reduced as $R_{ee}\sim \phi^{-1/2}$.  

Provided that a chain of length $N$ in 2D polymer solution is divided into $N/g$ blobs of size $\xi$, each comprised of $g$ \emph{correlated} monomers ($\xi\simeq ag^{\nu}$), the area fraction of the monomers inside the blob is $\phi\simeq ga^2/\xi^2$. 
From these two relations, it follows that  
\begin{align}
\xi(\phi)\simeq a\phi^{\frac{\nu}{1-2\nu}}
\label{eqn:corrlength}
\end{align} 
and 
\begin{align}
g(\phi)\simeq \phi^{\frac{1}{1-2\nu}}.
\end{align} 
Since $\nu=3/4$, $4/7>1/2$, 
the size and number of blobs decrease with $\phi$. 
With this blob picture in mind, the $\phi$-dependent size of polymer $R_{ee}(\phi)$ is expected to scale as 
\begin{align}
R_{ee}(\phi)\simeq \xi(\phi)(N/g(\phi))^{\nu}.
\label{eqn:Ree}
\end{align} 
Plotted by rescaling $R_{ee}(\phi)$ with $\xi(\phi)$ and $N$ with $g(\phi)$, 
the individual curves of $R_{ee}(N,\phi)$ obtained at varying $N$ and $\phi$ in Fig.~\ref{singConf}B collapse on the two distinct master curves (Fig.~\ref{singConf}C). 
(i) For the value of $\phi$ in which the blob size is greater than that of a chain ($R_{ee}/\xi(\phi)<1$, $N/g(\phi)<1$), $R_{ee}/\xi(\phi)\simeq (N/g(\phi))^{\nu}$ with $\nu=4/7$ for $\Theta$ solvent and $\nu=3/4$ for good solvent condition (Fig.~\ref{singConf}C). 
(ii) For the opposite case ($R_{ee}/\xi(\phi)>1$, $N/g(\phi)>1$), 
all the data points obtained from different $N$s and solvent qualities collapse onto the single master curve, $R_{ee}/\xi(\phi)\simeq (N/g(\phi))^{1/2}$.  
From (i) and (ii), it is suggested that the effect of solvent quality on the chain manifests itself only inside blobs, beyond which the individual polymers obey the statistics of polymer melts ($R_{ee}\sim N^{1/2}$). 

In fact, the blob size $\xi(\phi)$ is equivalent to the correlation length of the polymer solution. In semi-dilute phase, the correlation length $l(\phi)$ can be associated with the Flory radius as $l(\phi)\sim R_F(\phi^\ast/\phi)^{m_f}$ \cite{deGennesbook}. 
Since $\phi^\ast\sim N^{1-2\nu}$, $R_F\simeq aN^{\nu}$, and $l(\phi)$ should be independent of the chain length ($N$) of individual polymer in solution, one can determine $m_f$ from $\nu+m_f(1-2\nu)=0$. 
Thus, $l(\phi)\simeq a\phi^{\nu/(1-2\nu)}$, which is equivalent to Eq.~\ref{eqn:corrlength}, allowing to interpret that the blob size $\xi(\phi)$ is tantamount to the correlation length of the polymer solution in semi-dilute phase. 

When $g^{1/2}\simeq \phi^{1/2}\xi/a$ is substituted to Eq.~\ref{eqn:Ree} with $\nu=1/2$, 
it yields $R_{ee}\simeq aN^{1/2}\phi^{-1/2}$, which accounts for the $R_{ee}\sim\phi^{-1/2}$ ($\phi\gg \phi^\ast$) shown in Fig.~\ref{singConf}B. 

\begin{figure}[t]
\includegraphics[width=1.0\linewidth]{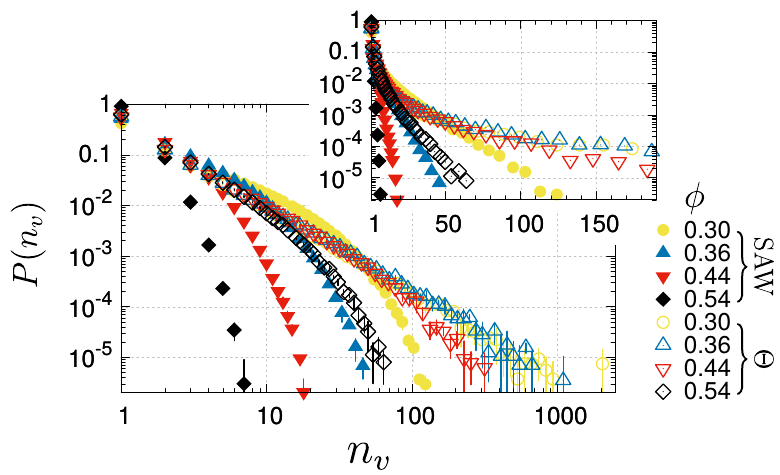}
\caption{
Probability distribution of void size ($n_{v}$) in semi-dilute polymer solutions ($\phi^\ast\leq\phi\ll 1$) with $N=640$.  
The inset shows the log-linear plot. 
The corresponding configurations of polymer solution are depicted in Figs.~2 and ~S1. 
}
\label{void}
\end{figure}

\begin{figure}[t]
\includegraphics[width=0.8\linewidth]{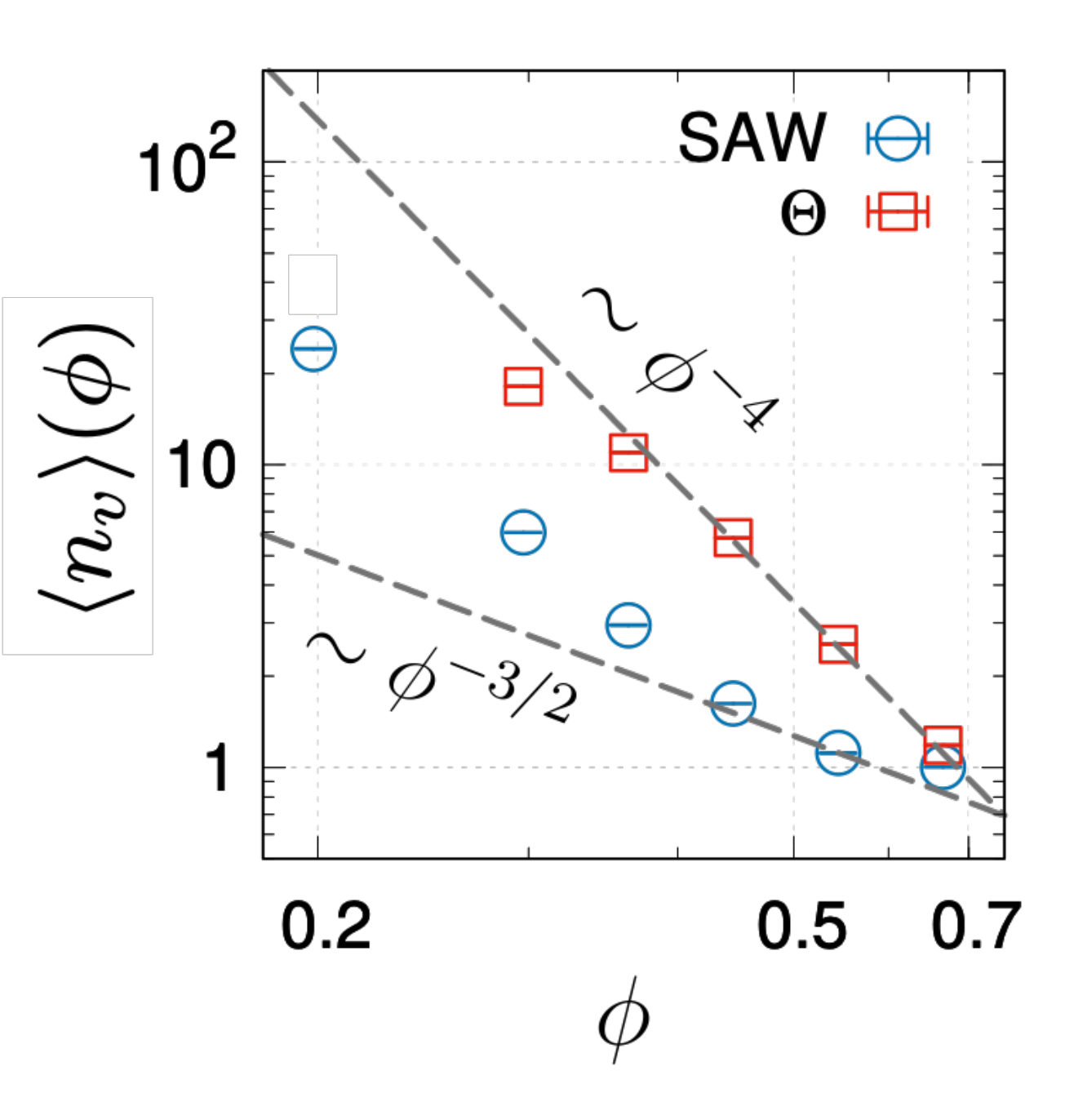}
\caption{The average size of interstitial voids, $\langle n_v\rangle$ under good and $\Theta$ solvent conditions are calculated from $P(n_v)$ in Fig.~\ref{void}. As expected, the void size is a decreasing function of $\phi$. For $\phi>\phi^\ast$ ($\phi^\ast_{\rm SAW}\approx 0.02$, $\phi^\ast_{\Theta}\approx 0.27$ for $N=640$) the void size scales with $\phi$ as $\langle n_v\rangle\sim \phi^{-4}$ and $\sim\phi^{-3/2}$ for $\Theta$ and SAW solution, respectively.}
\label{void_size}
\end{figure}

\subsection*{Interstitial voids} 
The varying sizes of interstitial voids interspersing the space between monomers in $\Theta$ condition are another key feature that differentiates $\Theta$ polymer solution from SAW solution in 2D at the same $\phi$ (Fig.~\ref{conf1}C, see also Figs.~\ref{conf_S1} and S2 calculated with $N=640$ for clearer images).  
In comparison with the interstitial voids in $\Theta$ chain solution, those in SAW solution appear more uniform in size. 


To make this observation more quantitative, we first identify the voids from the configurations of polymer solution and calculate the void size distribution (Fig.~\ref{void}). 
The whole simulation box was divided into cells of $1.5 a\times 1.5 a$ square lattice, and a void was defined as a cluster of empty cells that are connected without being intercepted by the polymer chains. 
The Hoshen-Kopelman algorithm, which is often utilized in studies of percolation~\cite{hoshen1976PRB,Binder2010Book}, was employed to quantify the size of a void by means of the number of unoccupied cells ($n_{v}$). 
The void size distributions $P(n_v)$ in Fig.~\ref{void} have exponentially decaying tails. 
The tail of $P(n_v)$ for the $\Theta$ chain solution at the same $\phi$ is an order of magnitude longer than that for the SAW solution. 

Notably, the interstitial voids formed in $\Theta$ chain solution display {\color{black}a} more heterogeneous distribution with heavier tails, and hence the average void size $\langle n_v\rangle\left(=\int n_vP(n_v)dn_v\right)$, which can be calculated from Fig.~\ref{void}, is greater than that of SAW chain solution, i.e., $\langle n_v^{\Theta}\rangle(\phi) > \langle n_v^{\rm SAW}\rangle(\phi)$ (Fig.~\ref{void_size}). 
Larger interstitial voids in polymer solution alleviate the inter-monomer repulsion, lowering the osmotic pressure. 
From the analysis of our numerics, we find that the average size of the interstitial void scales approximately with $\phi$ as 
$\langle n_v\rangle \sim \phi^{-4}$ for $\Theta$ chain solution and $\langle n_v\rangle \sim \phi^{-3/2}$ for SAW solution for $\phi>\phi^\ast$ (Fig.~\ref{void_size}). 
Notably, the average size of interstitial void displays the scaling relation identical to 
that of the blob size (or correlation length). 
In fact, the surface pressure in semi-dilute phase is related with $\xi(\phi)$ as 
\begin{align}
\Pi\sim \frac{k_BT}{\xi(\phi)^2}\sim \frac{k_BT}{\phi^{\frac{2\nu}{1-2\nu}}}.   
\label{eqn:pressure_xi}
\end{align}
Due to Eq.\ref{eqn:pressure_xi}, the inequality of $\xi_\Theta(\phi) > \xi_{\rm SAW}(\phi)$ (or $\langle n_v^{\Theta}\rangle(\phi) > \langle n_v^{\rm SAW}\rangle(\phi)$) for $0<\phi<1$ implies the inequality of 
$\Pi_\Theta(\phi)<\Pi_{\rm SAW}(\phi)$. 
For the case of dilute solution ($\phi<\phi^\ast$), 
$\Pi\sim (\phi/N)+B_2(\phi/N)^2+B_3(\phi/N)^3+\cdots$ (see Appendix A). 
Since $B_2>0$ and $B_3>0$ for a SAW chain and $B_2(\phi/N)^2+B_3(\phi/N)^3+\cdots\approx 0$ for a $\Theta$ chain, 
the inequality $\Pi_\Theta(\phi)<\Pi_{\rm SAW}(\phi)$ is expected as well.

\section{Concluding Remarks} 
Our numerics, {\color{black}although the chains are not still long enough to discuss the scaling regime}, have {\color{black}semi-quantitatively} reproduced the basic features characterizing the experimentally measured $\Pi$-$\phi$ isotherm of thin polymer film \cite{vilanove1980PRL,witte2010macromolecules}. 
Polymer configurations of 2D polymer solution visualized through our numerics clarify a qualitative difference between the polymer configurations in good and $\Theta$ solvents. 

Among the two fundamental scaling exponents involved in polymer configurations, $\nu$ and $\gamma$, the one involved with the size ($R\sim N^{\nu}$) and the other with the entropy of the chain ($Z_N\sim \mu^NN^{\gamma-1}$) \cite{deGennesbook}, the $\nu$ is the only exponent that decides the dependence of the osmotic pressure on $\phi$.  
{\color{black}It is worth noting that} the exponent $\gamma$, (more specifically $\gamma_4$ and $\gamma_2$, where $\gamma_L$ is the exponent for $L$-star polymer \cite{Duplantier89JSP}) which may be linked to the correlation hole exponent $\theta$ and the fractal dimension of external perimeter $d_p$ \cite{deGennesbook,Duplantier89JSP}, however,  
make no contribution to determining the osmotic pressure. 

{\color{black}Visualizing the distributions of monomers (Figs.~\ref{conf1}, \ref{conf_S1}, and \ref{fig:gr}) and more importantly the distinct distributions of interstitial void for two different polymer solutions (Figs.~\ref{void} and \ref{void_size}), this study offers comprehensive understanding to the physical origin of the differing surface pressure isotherms of 2D polymer solution under good and $\Theta$ solvent conditions.} \\

\section{Methods}
\subsection*{Generating $\Theta$ chains in two dimensions.}  
\begin{figure}[t]
\includegraphics[width=1.0\linewidth]{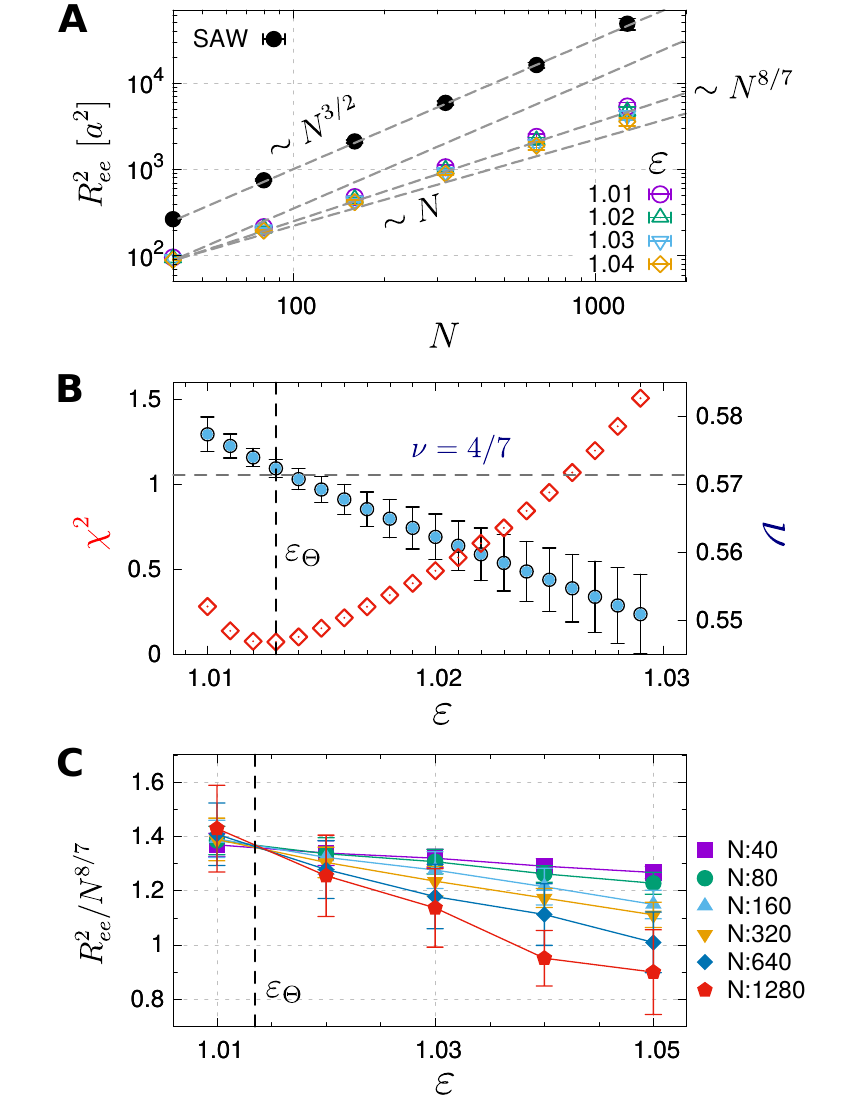}
\caption{
Single chain size scaling and the $\Theta$ condition in 2D. 
(A) While the end-to-end distance $R_{ee}$ of SAW chain shows the expected scaling behavior ($R^{2}_{ee} \sim N^{3/2}$), 
the monomer attraction strength ($\varepsilon$) were fine-tuned such that 
$R_{ee}(N,\epsilon)$ satisfies a power law $R^{2}_{ee} \sim N^{2 \nu}$ with some value of $\nu$. 
(B) $\chi^{2}$ and the best value of $\nu$ obtained by fitting the data at different values of $\varepsilon$ to the power law. 
The most confident scaling relation with $\nu \simeq 4/7$ 
is acquired when $\varepsilon_{\Theta} \simeq 1.013$. 
(C) Alternatively, if we take Duplantier's proposal $\nu=4/7$ as granted, $R^{2}_{ee}/N^{2\nu}$ should become independent of $N$ at the $\Theta$ condition, 
which again locates $\varepsilon_{\Theta}$ around 1.013 in our model.}
\label{sigchain}
\end{figure}

The following energy potential was used to simulate a polymer chain composed of $N$ segments.
\begin{align}
\mathcal{H}({\bf r}) = \mathcal{H}_{b}({\bf r}) + \mathcal{H}_{nb}({\bf r}), 
\label{eqn:potential}
\end{align}
where ${\bf r} = \{ {\bf r}_{i} \}$ and ${\bf r}_{i}$ denotes the coordinate of the $i$-th monomer in a 2D plane. 
The first term models the chain connectivity with the finite extensible nonlinear elastic (FENE) potential and a shifted Weeks-Chandler-Anderson (WCA) potential, 
\begin{widetext}
\begin{equation}
\label{hb}
\beta\mathcal{H}_{b}({\bf r}) = -\frac{\beta k}{2} R^{2}_{c} \sum_{i=0}^{N-1} \log\left(1-\frac{r_{i,i+1}^{2}}{R^2_c}\right) + \sum_{i=1}^{N} 4\left[\left(\frac{a}{r_{i,i+1}}\right)^{12} - \left(\frac{a}{r_{i,i+1}}\right)^{6} + \frac{1}{4}\right] H(2^{1/6}a - r_{i,i+1}), 
\end{equation}
\end{widetext}
where $r_{i,i+1} \equiv |{\bf r}_{i+1} -{\bf r}_{i}|$ is the segment length, $H(\cdots)$ is the Heaviside step function, and we chose the parameters $k=30~k_\text{B}T$ with $R_{c} = 1.5$ $a$. 
The energy potential with these parameters equilibrates the segments at $b_i\approx a$. 
The second term in Eq.\ref{eqn:potential} involves 
the non-bonded interactions between two different monomers. 
For good solvent $\mathcal{H}_{nb}({\bf r})=\mathcal{H}_{nb}^\text{good}({\bf r})$
\begin{widetext}
\begin{align}
\label{hnb}
\beta\mathcal{H}_{nb}^\text{good}({\bf r}) = 
  \sum_{i<j} 4\left[\left(\frac{a}{r_{ij}}\right)^{12} - \left(\frac{a}{r_{ij}}\right)^{6} + \frac{1}{4}\right] H(2^{1/6}a - r_{ij}),
  \end{align}
  and for $\Theta$ condition $\mathcal{H}_{nb}({\bf r})=\mathcal{H}_{nb}^\Theta({\bf r})$
 \begin{align}
\beta\mathcal{H}_{nb}^{\Theta}({\bf r}) =    \sum_{i<j} \varepsilon\left[\left(\frac{a}{r_{ij}}\right)^{12} - 2\left(\frac{a}{r_{ij}}\right)^{6} + \Delta_s\right] H(2.5a - r_{ij}), 
\end{align}
\end{widetext}
where $r_{ij} = |{\bf r}_i - {\bf r}_j|$. 
For the $\Theta$ solvent, the Lennard-Jones potential was shifted upward by $\Delta_{s} = 2\times0.4^{6} - 0.4^{12}$ such that the potential is continuous at $r_{ij}=2.5$ $a$, 
and the parameter $\varepsilon$ was set to $\varepsilon_{\Theta} = 1.013$ which yields the scaling $R_{ee} \sim N^{4/7}$ (see Fig.~\ref{sigchain}).

To sample polymer solution configuations, 
we integrated the underdamped Langevin equations:
\begin{equation}
m {\ddot {\bf r}}_{i} = -\zeta {\dot {\bf r}}_{i} - \nabla_{{\bf r}_i} \mathcal{H}({\bf r}) + {\bf \xi}_{i}(t),
\end{equation}
with the random force satisfying $\langle {\bf \xi}_{i}(t)\rangle=0$ and $\langle{\bf \xi}_{i}(t) \cdot {\bf \xi}_{j}(t') \rangle=4 \zeta k_{B}T \delta_{ij}\delta(t-t')$. 
A small time step $\delta t = 0.005 \tau$ and a small friction coefficient $\zeta = 0.1 m/\tau$ with the characteristic time scale $\tau = \left(m a^{2}/\varepsilon\right)^{1/2}$ were employed to enhance the rate of equilibrium sampling of polymer configruations.

The polymer solutions of long chains ($N=80$, 160, 320, 640, 1280) were simulated in an NVT ensemble in two steps. 
(i) From a condition of dilute solution ($\phi = \pi/400\approx 7.85\times 10^{-3}$) that contains 36 pre-equilibrated chains, 
the size of the periodic box was reduced step by step with $L\rightarrow \eta L$ ($\eta = 0.904$), so that 
the area fraction is increased by a factor of $\eta^{-2}$ in each step.  
At each value of $\phi$, excessive shrinking-induced overlaps between monomers were eliminated by gradually increasing the short-range repulsion part of $\mathcal{H}$. 
More specifically, the non-bonded potential $\mathcal{H}_{nb}({\bf r})$ was replaced with $\min\{u_{c}, \mathcal{H}_{nb}({\bf r})\}$, in which $u_{c}$ was slowly elevated. 
(ii) For the production run, the system was simulated for $500N\tau$, and chain configurations were collected every $0.1N\tau$. 
For each combination of $N$ and $\phi$, 10 replicas were generated from different initial configurations and random seeds. 

To facilitate the sampling to calculate $\Pi$ more accurately, polymer solutions of short chains ($N=40$, 50, 56, 70) 
were simulated in an isothermal-isobaric (NPT) ensemble, in which the total number of monomers was fixed to 8400.
During the production run for $1.5\times 10^{4}N\tau$, area-changing trial moves were generated every $N\tau$ time step via the Metropolis algorithm, which helped maintain the system at a constant pressure \cite{2001Frenkel}. 
The structural properties averaged over all replicas were demonstrated in this study with the error bars denoting the standard deviations. 
The simulations were performed using the ESPResSo 3.3.1 package \cite{limbach2006espresso}.
\\

\subsection*{Standard isothermal compressibility $\kappa_\text{T}$.} 
As simulations of short chain solutions were performed in an isothermal-isobaric (NPT) ensemble, 
we calculated the standard isothermal compressibility by 
\begin{equation}
\kappa_\text{T} = \frac{1}{k_\text{B}T} \frac{\left<A^2\right> - \left<A\right>^{2}}{\left<A\right>}, 
\end{equation}
where $A$ is the fluctuating area of the simulated solution \cite{Hansen1990Book}. 

For longer chains, $\kappa_\text{T}$ was calculated by $\kappa_\text{T} =  \chi_\text{T} (\rho k_\text{B}T)^{-1}$, 
where the reduced isothermal compressibility $\chi_\text{T}$ was determined with a spatial block analysis method \cite{heidari2018Entropy}. 
More specifically, whereas $\chi_\text{T}$ can be calculated by 
\begin{equation}
\chi_\text{T} = \frac{\langle N^2\rangle - \langle N\rangle^{2}}{\langle N\rangle} 
\end{equation}
where $N$ is the particle number in a grand ensemble, 
various finite-size effects need to be considered to extrapolate $\chi_\text{T}$ in a simulated canonical (NVT) ensemble. 
$\chi^{B}_\text{T}$ calculated from subdomains in a NVT ensemble varies with the domain size $B$ (Fig.~S3A). 
When the ratio of subdomain size $\lambda \equiv B/B_{0}$ approaches 1, where $B_0$ denotes the size of the full simulation box, $\chi_\text{T}$ approaches 0 as expected. 
Next, we extrapolated the values of $\chi^{\infty}_\text{T}$ by fitting the data to 
a function proposed in Ref.~\cite{heidari2018Entropy}, $\chi^{B}_\text{T} = \chi^{\infty}_\text{T} \lambda (1-\lambda^3) - c$ (Fig.~S3B). 
The final results are plotted as a function of the area fraction in Fig.~\ref{kappaT}B.
\\

\noindent{\bf Acknowledgement. }
We thank Prof. Bae-Yeun Ha for a number of useful comments. 
This work is in part supported by National Natural Science Foundation of China, ZSTU intramural grant (12104404, 20062226-Y to L.L.), and KIAS Individual Grant (CG035003 to C.H.) at Korea Institute for Advanced Study.
We thank the Center for Advanced Computation in KIAS for providing computing resources.


\section{Appendix} 
\setcounter{equation}{0}
\setcounter{figure}{0}

\renewcommand{\theequation}{A\arabic{equation}}
\renewcommand{\thefigure}{A\arabic{figure}} 

\renewcommand{\theequation}{A\arabic{equation}}

\subsection{Osmotic pressure of polymer solution with increasing $\phi$}

\begin{figure}[t]
        \includegraphics[width=1.0\linewidth]{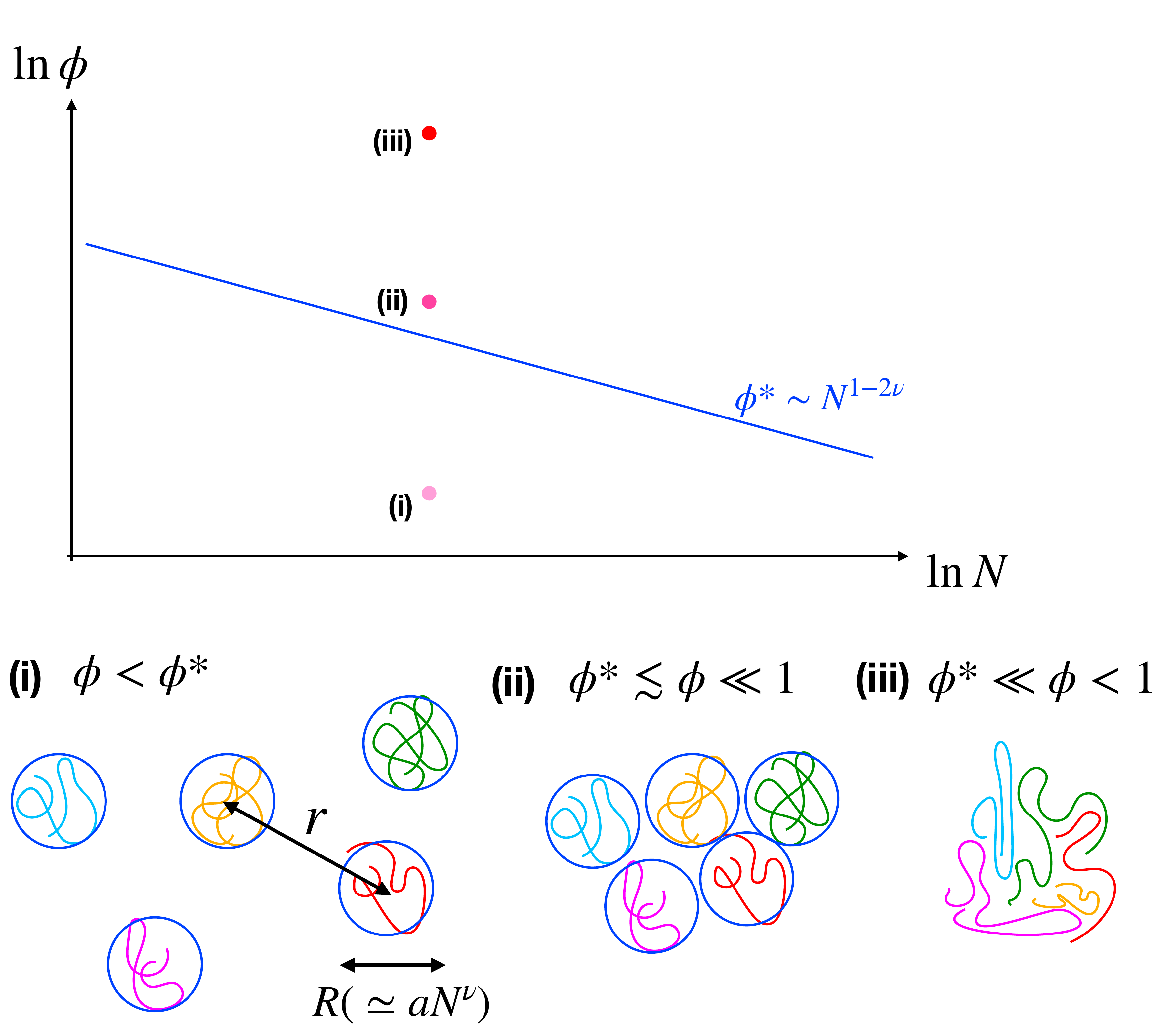}
		\caption{Diagram illustrating the polymer solution at three different regimes of area fraction. 
		In isolation, the size of polymer scales with $N$ as $R\simeq aN^{\nu}$. 
	The average distance between polymers is denoted by $r$. 
		The threshold overlap area fraction ($\phi^\ast$, blue line) scales with the length of polymer chain as $\phi^\ast\sim N^{1-2\nu}$ in 2D ($d=2$). 
		Depending on the value of area fraction $\phi$, the polymer solution is classified into three regimes: 
		(i) dilute ($\phi<\phi^\ast$), (ii) semi-dilute ($\phi^\ast < \phi \ll 1$), and (iii) concentrated regime ($\phi^\ast\ll \phi < 1)$. 
		\label{diagram}}
\end{figure}

The osmotic pressure of polymer solution displays substantial changes with increasing $\phi$. 
Before our in-depth discussion on the solvent quality dependent osmotic pressure, we briefly review some basics of polymer solution along with Fig.~\ref{diagram}.

(i) In non-overlapping dilute regime ($\phi<\phi^\ast$) 
each chain is effectively isolated and the property of polymer solution can be described with individual polymer chains that behave like a van der Waals gas of radius $R_F\sim N^{\nu}$ at concentration $\sim \phi/N$. 
In this regime, the osmotic pressure is given by $a^d\Pi/T\simeq \phi/N+B_2(\phi/N)^2+\mathcal{O}[(\phi/N)^3]$, where $a$ is the monomer size and $B_2\sim R_F^d$ \cite{deGennesbook}.

(ii) As $\phi$ increases, there is a point where 
the average distance between the chains and their size ($R_F$) becomes comparable, and the polymer solution reaches the overlap concentration ($\rho\approx \rho^\ast$ or $\phi\approx\phi^\ast$). 
Beyond this point, it becomes difficult to tell whether neighboring monomers belong to the same chain or to the different chain, and the global concentration of monomers becomes identical to the intrachain monomer concentration $\rho^\ast=N/R_F^d$. 
Hence the corresponding volume fraction is given as 
$\phi^\ast=\rho^\ast a^d=Na^d/R_F^d\sim N^{1-d\nu}$. 
For polymer solution in the regime of semi-dilute condition, $\phi^\ast< \phi\ll 1$,   
its osmotic pressure obeys the scaling law of $a^d\Pi/T\simeq (\phi/N)f(\phi/\phi^\ast)$. 
$\Pi$ in this regime is impervious to the actual length of polymer ($N$), and it is determined solely by the local volume (area) fraction $\phi$ \cite{vilanove1980PRL,witte2010macromolecules,deGennesbook}. 
The scaling ansatz that the scaling function is given by $f(x)\sim x^m$ results in 
$(\phi/N)(\phi/\phi^\ast)^m\simeq \phi^{1+m}N^{-1+m(d\nu-1)}\sim N^0$, which determines $m=(d\nu-1)^{-1}$ and yields the scaling relation $\Pi\sim \phi^{\frac{d\nu}{d\nu-1}}$.    

(iii) When $\phi$ increases further, the pressure of polymer solution starts to deviate from $\Pi\sim \phi^{\frac{d\nu}{d\nu-1}}$ \cite{paturej2019PRL}, and in highly concentrated regime ($\phi\gg \phi^\ast$) the solution eventually forms a polymer melt.  
In 3D ($d=3$), the interaction between monomers of polymer chains are effectively screened, so that the chains behave like ideal polymers with the size of individual polymer chains scaling as $R_F (=\langle R_g^2\rangle^{1/2}) \sim N^{1/2}$.  
The polymer chains \emph{interpenetrate} each other, displaying 
a strong correlation with its neighbors \cite{deGennesbook}.  
By contrast, polymer melts in 2D are characterized by completely different physical properties because of the topological interaction overwhelming other interactions. 
Polymer chains \emph{segregate} from each other, and form compact, space-filling domains whose size scales as $R_F\sim N^{1/d}$ with $d=2$.  
It is of particular note that although the scaling exponents in the melts are identical to $\nu=1/2$ for both 2D and 3D, the underlying physics giving rise to the exponent $1/2$ are fundamentally different \cite{schulmann2013PSSC}.  
In 2D, the intrachain monomer distributions of polymer chain in polymer solution under both good and $\Theta$ solvent conditions differ from the distribution of Gaussian polymer (Fig.~S4) \cite{beckrich2007macro}. \\

\setcounter{equation}{0}

\renewcommand{\theequation}{B\arabic{equation}}

\subsection{Connection between $\kappa_T$, $g(r)$, and number fluctuations} 
For $N$ indistinguishable particles distributed in space, a grand partition function is written as 
\begin{align}
\Xi=\frac{1}{N!}\sum_{N=1}^{\infty}z^N\int d{\bf r}^Ne^{-\beta V_N}. 
\end{align}
where $z\equiv e^{\beta\mu}$, $d{\bf r}^N\equiv \prod_{i=1}^N d{\bf r}_i$, $V_N\equiv V({\bf r}_1,{\bf r}_2,\ldots {\bf r}_N)$, and the factor $N!$ is introduced to account for the indistinguishability of the particles. 
Then the joint probability density of $n$ indistinguishable particles in space is given 
\begin{widetext}
\begin{align}
\rho^{(n)}({\bf r}_1,{\bf r}_2,\ldots,{\bf r}_n)&=\frac{1}{\Xi}\left[\frac{1}{(N-n)!}\sum_{N>n}^{\infty}z^N\int d{\bf r}^{N-n}e^{-\beta V_N}\right]=\frac{N!}{(N-n)!}\underbrace{\frac{\sum_{N>n}^{\infty}z^N\int d{\bf r}^{N-n}e^{-\beta V_N}}{\sum_{N=1}^{\infty}z^N\int d{\bf r}^Ne^{-\beta V_N}}}_{\equiv P({\bf r}_1,{\bf r}_2,\ldots,{\bf r}_n)},
\label{eqn:rho_n} 
\end{align}
\end{widetext}
where $P({\bf r}_1,{\bf r}_2,\ldots,{\bf r}_n)$ is the joint probability density of $n$ \emph{distinguishable} particles. Then it follows from Eq.\ref{eqn:rho_n} that 
\begin{align}
\int d{\bf r}^n\rho^{(n)}({\bf r}_1,{\bf r}_2,\ldots,{\bf r}_n)=\Big\langle\frac{N!}{(N-n)!}\Big\rangle. 
\end{align}

Now, we consider the joint probability density of two indistinguishable particles at ${\bf r}_1$ and ${\bf r}_2$ with an assumption that their distribution in space is \emph{homogeneous} and \emph{isotropic}, satisfying 
$\rho^{(2)}({\bf r}_1,{\bf r}_2)\approx \rho^2g(r_{12})$ where $\rho\equiv N/V$, $r_{12}\equiv |{\bf r}_1-{\bf r}_2|$ and $g(r)$ is the radial distribution function. 
Then, together with the probability density for a single particle in homogeneous and isotropic space, satisfying $\rho({\bf r})=\rho=N/V$ for $d=3$ (or $\rho=N/A$ for $d=2$), we obtain  
\begin{widetext}
\begin{align}
\int\int[\rho^{(2)}({\bf r}_1,{\bf r}_2)-\rho^{(1)}({\bf r}_1)\rho^{(1)}({\bf r}_2)]d{\bf r}_1d{\bf r}_2
&=\langle N\rangle\rho\int d{\bf r}[g({\bf r})-1]=\langle N^2\rangle-\langle N\rangle-\langle N\rangle^2, 
\end{align}
\end{widetext}
which yields $\langle (\delta N)^2\rangle/\langle N\rangle=1+\rho\int [g({\bf r})-1]d{\bf r}$ (Eq.\ref{eqn:gr_kappa}). 

Next, the total differential $d\Xi=-SdT+Ad\Pi-Nd\mu$ of the grand ensemble $\Xi=\Xi(T,\Pi,\mu)$ gives the following relations at constant temperature, 

\begin{align}
N\left(\frac{\partial \beta\mu}{dN}\right)_{T,A}&=\beta A\left(\frac{\partial \Pi}{dN}\right)_{T,A}\nonumber\\
&=-\beta A\left(\frac{\partial \Pi}{\partial A}\right)_{T,N}\left(\frac{\partial A}{\partial N}\right)_{T,\Pi}\nonumber\\
&=\frac{1}{\rho k_BT\kappa_T}
\label{eqn:N_fluctuations}
\end{align}
with $\rho\equiv N/A$. 
Together with $(\partial\log{\Xi}/\partial\beta\mu)_{T,A}=\langle N\rangle$ 
and $(\partial\langle N\rangle/\partial\beta\mu)_{T,A}=\langle(\delta N)^2\rangle$, it follows from Eq.~\ref{eqn:N_fluctuations} that 
\begin{align}
\frac{\langle (\delta N)^2\rangle}{\langle N\rangle}=\rho k_BT\kappa_T. 
\end{align}


%

\clearpage 

\setcounter{equation}{0}
\setcounter{figure}{0}
\setcounter{table}{0}
\renewcommand{\theequation}{S\arabic{equation}}
\renewcommand{\thefigure}{S\arabic{figure}} 
\renewcommand{\thetable}{S\arabic{table}}

\section*{Supplementary Information}

\subsection*{Scattering functions to probe the intra-chain configurations of a polymer} 
When $g(r)$ is decomposed into intra- and interchain contributions ($g_\text{intra}(r)$ and $g_\text{inter}(r)$), 
the qualitative difference between the two solvent conditions  becomes clear in $g_\text{inter}(r)$ (the lower panels of Fig.~4B). 
Although the difference between $g_\text{intra}^{\rm SAW}(r)$ and $g_\text{intra}^{\Theta}(r)$ in real space is not clear, 
the intrachain configurations under the two solvent conditions can be differentiated by means of 
their Fourier transformed version (see Fig.~\ref{intraFxy}), namely the intra-chain form factor $F(k)$ (or the Kratky plot, $k^2F(k)$ versus $k$) in the range from the dilute to semi-dilute condition ($\phi < 0.44$). 
Neutron scattering experiment on \emph{deuterated} single polymer chain in solution can, in practice, be employed to measure the intrachain form factor.   

(i) In semi-dilute regime, for the intermediate range of wavevector $k$ ($R_F^{-1}<k<\xi(\phi)^{-1}$), which is used to probe the local region of chain structure, 
the number density of monomer $\rho({\bf r})\sim n(r)/r^d$ with $\int \rho({\bf r})d{\bf r}=N$ 
can be Fourier transformed to define the form factor $F(k)$ as 
$F(k)\sim \int \rho({\bf r})e^{i{\bf k}\cdot{\bf r}}d{\bf r}$. 
By performing the dimensional analysis with the relations $r\sim an^{\nu}$ and $d{\bf r}\sim r^{d-1}dr$, 
it is straightforward to show the following relationship of $F(k)$,  
\begin{align}
F(k)&\sim k^{-1/\nu}. 
\end{align}
This yields $F(k)\sim k^{-4/3}$ for SAW and $F(k)\sim k^{-7/4}$ for $\Theta$ polymer in polymer solution. 

(ii) In concentrated regime, the chains are compact. 
In this case, the scattering amplitude is mainly contributed by the monomer exposed to the surface (perimeter) of the domain formed by the chain, giving rise to the Porod scattering \cite{debye1957scattering,bale1984PRL,wong1988PRL}. 
\begin{align}
F(k)\sim k^{-(2d-d_p)}=k^{-(d+\theta)}
\end{align}
where $d$ is the dimensionality, $d_p(=d-\theta)$ is the fractal dimension of the surface (perimeter), and $\theta=\theta_2$ is Duplantier's contact exponent of chain in 2D.  For compact chain in 2D at $\phi\gg \phi^\ast$, 
there is no distinction between the polymer configurations in the two solvent qualities, and hence the exponents of $d=2$, $\theta_2=3/4$, and $d_p=5/4$ lead to $F(k)\sim k^{-11/4}$ and $k^2F(k)\sim k^{-3/4}$ for $\phi\approx 0.67$ (Fig.~\ref{intraFxy}).

\clearpage

\begin{figure*}[t]
\includegraphics[width=1\linewidth]{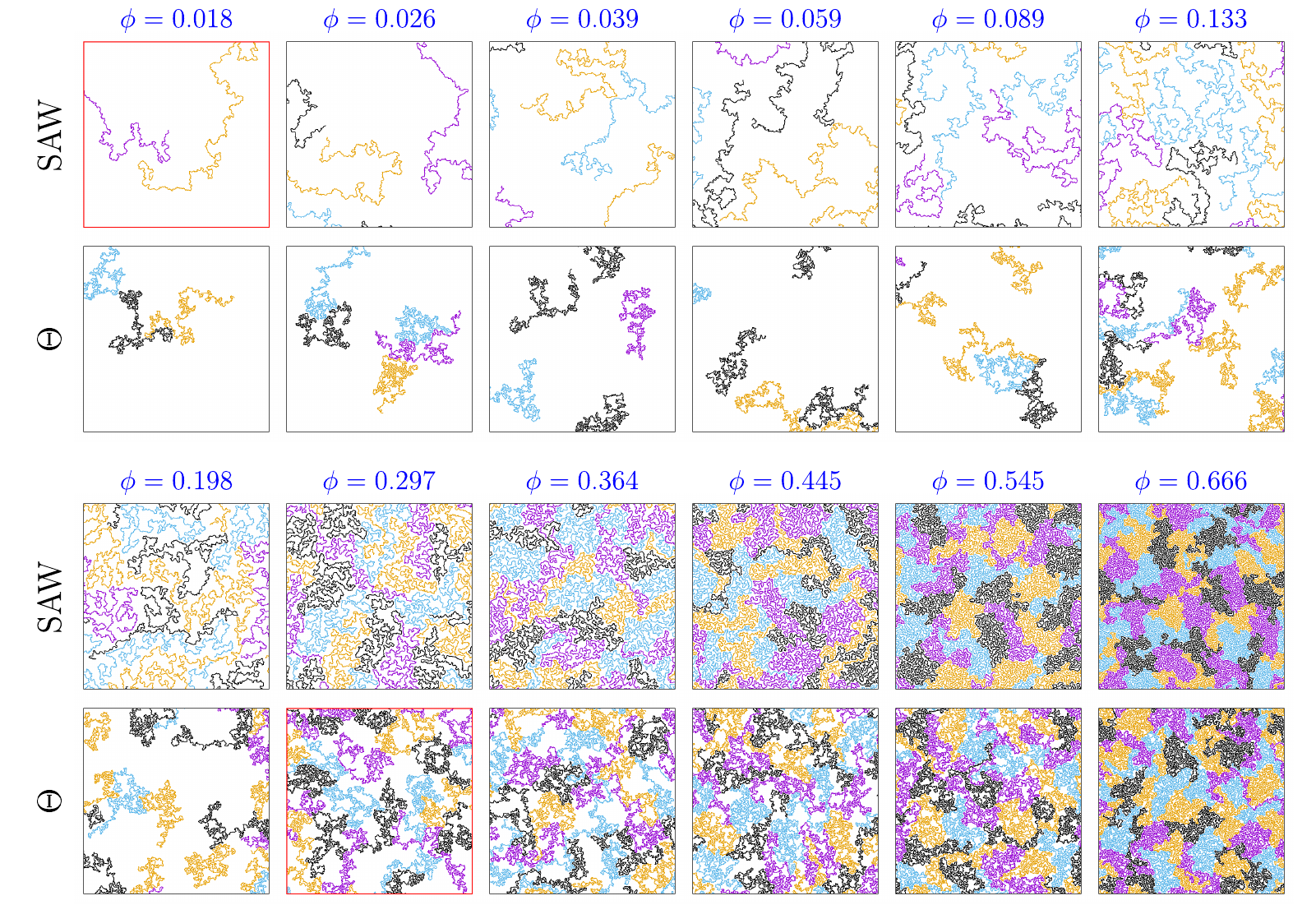}
\caption{
Same with Fig.~2, but shown are the configurations of polymer solutions in the 2D box at the identical field of view. 
As a result, only a part of the simulation is depicted in each panel except for the case with $\phi = 0.666$. 
}
\label{conf_S2}
\end{figure*}

\begin{figure*}[h!]
\includegraphics[width=0.7\linewidth]{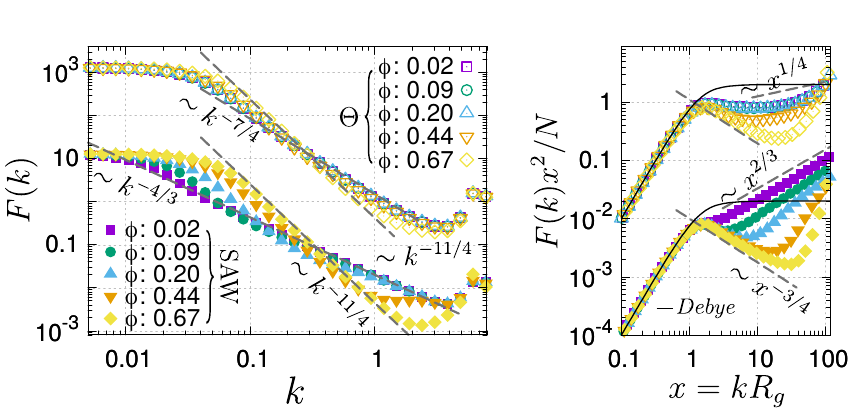}
\caption{
Intrachain form factor. 
(A) $F(k)$ calculated for individual chains ($N=1280$) in polymer solutions under good and $\Theta$ solvent conditions with varying $\phi$'s. 
For the sake of visual clarity, the plots of $F(k)$ for the two different solvent conditions are vertically shifted and separated. 
(B) Kratky plot of the same data using a rescaled wavevector $x=k R_{g}$. 
The Debye function for ideal chain, $F(k)/N = f_\text{D}(x^2)$ with $f_\text{D}(x) = 2(e^{-x}-1+x)/x^{2}$, is plotted using black solid lines for comparison.
}
\label{intraFxy}
\end{figure*}


\begin{figure*}[t]
\includegraphics[width=0.7\linewidth]{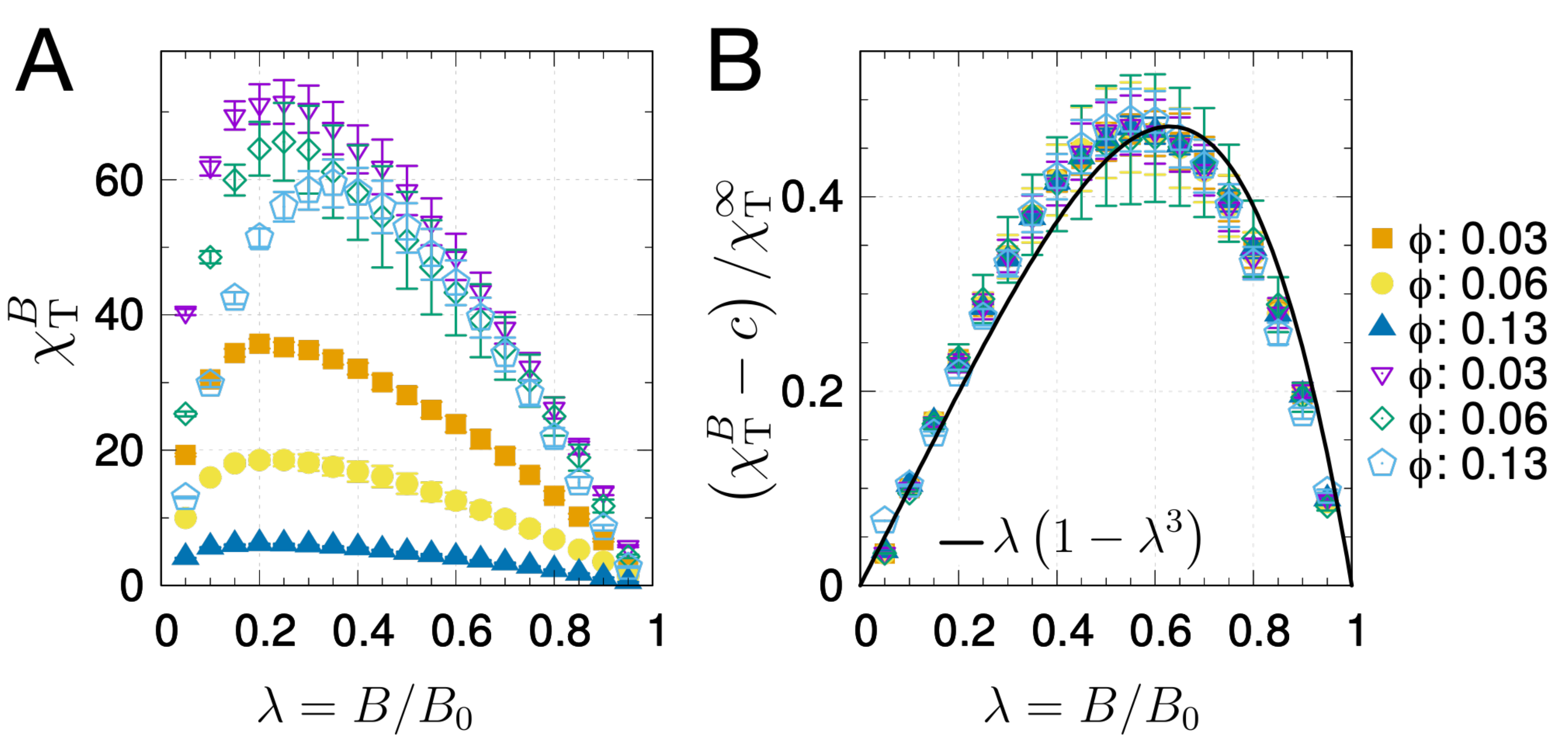}
\caption{
Spatial block analysis to compute the reduced isothermal compressibility $\chi_\text{T}$. 
Whereas $\chi_\text{T}$ can be calculated with $\chi_\text{T} = \left(\langle N^2\rangle - \langle N\rangle^{2}\right)/\langle N\rangle$ 
various finite-size effects need to be considered to extrapolate $\chi_\text{T}$ in simulations based on canonical (NVT) ensemble \cite{heidari2018Entropy}. 
(A) Typical $\chi^{B}_\text{T}$ calculated from square subdomains in an NVT ensemble of SAW (solid) and $\Theta$ chain (empty) solutions with $N=70$. 
The subdomain size is characterized by $\lambda = B/B_{0}$ where $B$ and $B_{0}$ are the sizes of the subdomains and the full simulation box, respectively. 
As $\lambda \rightarrow 1$, $\chi_\text{T}$ converges to $0$ in the NVT system. 
(B) Fitting data to the proposed relation, $\chi^{B}_\text{T} = \chi^{\infty}_\text{T} \lambda (1-\lambda^3) - c $, to extrapolate $\chi^{\infty}_\text{T}$ \cite{heidari2018Entropy}. 
The latter was used to calculate the standard isothermal compressibility shown in Fig.~3B by $\kappa_\text{T} = \chi^{\infty}_\text{T}/\rho k_\text{B}T$.
}
\label{chiT}
\end{figure*}

\begin{figure*}
\includegraphics[width=0.8\linewidth]{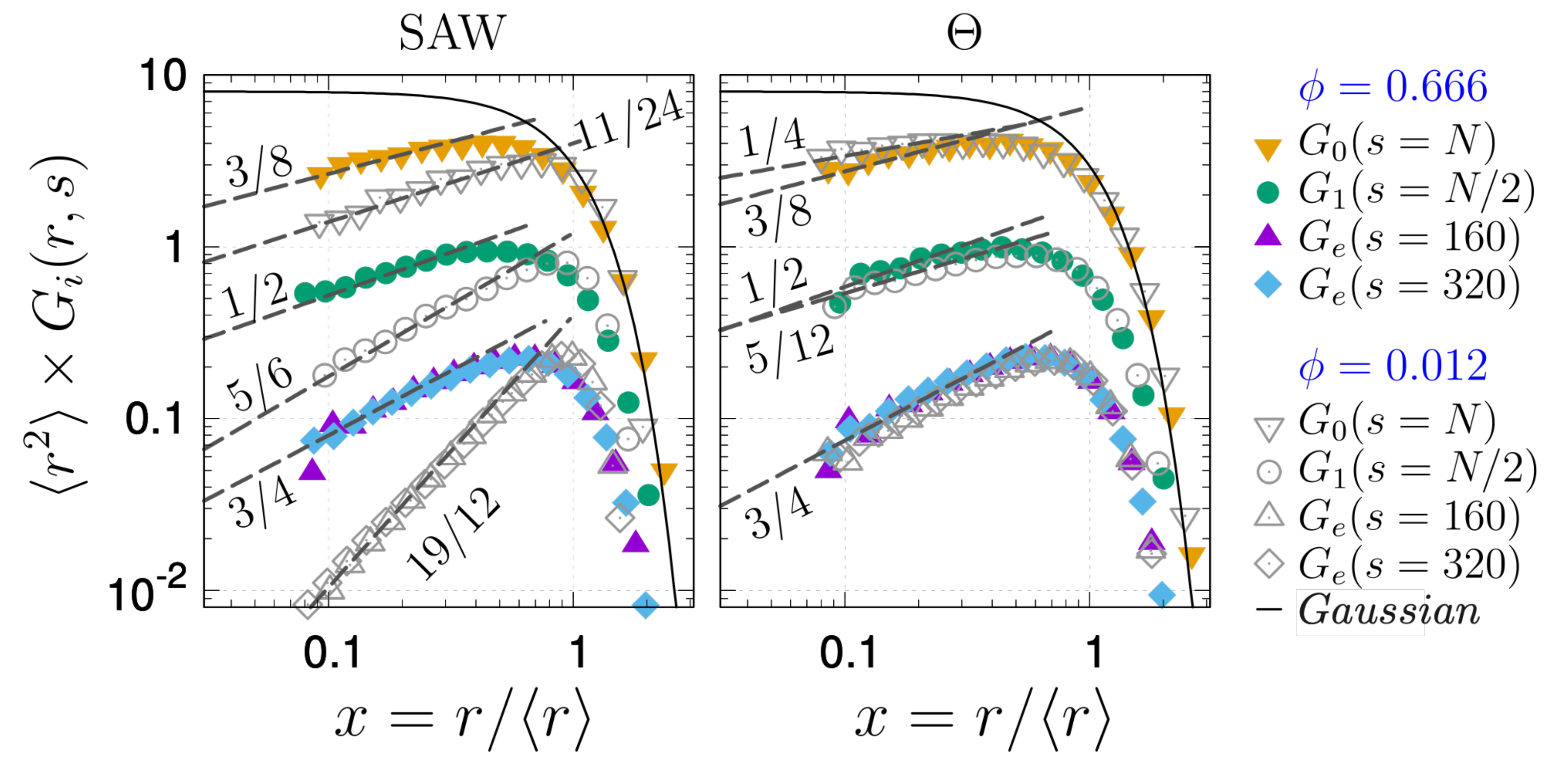}
\caption{Comparison between the intrachain monomer distribution $G_{i}(r,s)$ in the polymer solutions of SAW and $\Theta$ chain, 
at low ($\phi = 0.012$) and high ($\phi=0.667$) densities. 
$G_{0}(r,s=N)$ characterizes the chain end-to-end vector, 
$G_{1}(r,s=N/2)$ measures the distance between one end and the middle of a chain, 
and $G_{e}(r,s)$ averages the end-to-end distance of all subchains with a given contour length of $s$. 
$G_{e}(s=320)$ was calculated by using polymer chains with $N=1280$, and other three distributions were computed with $N=640$. 
The numbers next to the dashed lines denote the contact exponents predicted by Duplantier \cite{Duplantier87PRL,Duplantier89JSP}. 
Related discussions can be found in Fig.~9 of Ref.~\cite{schulmann2013PSSC}. 
The Gaussian distribution is plotted using black solid lines for comparison. 
}
\label{intraGs}
\end{figure*}

\end{document}